\documentclass[twocolumn,rsi,showpacs,preprintnumbers,amsmath,amssymb, superscriptaddress]{revtex4-1}
\usepackage{graphicx}% Include figure files
\usepackage{dcolumn}% Align table columns on decimal point
\usepackage{bm}% bold math
\usepackage{color}
\begin{document}

\preprint{}

\title{A precise method for visualizing dispersive features in image plots}
\author{P. Zhang}
\affiliation{Beijing National Laboratory for Condensed Matter Physics, and Institute of Physics, Chinese Academy of Sciences, Beijing 100190, China}

\author{P. Richard}
\affiliation{Beijing National Laboratory for Condensed Matter Physics, and Institute of Physics, Chinese Academy of Sciences, Beijing 100190, China}

\author{T. Qian}
\affiliation{Beijing National Laboratory for Condensed Matter Physics, and Institute of Physics, Chinese Academy of Sciences, Beijing 100190, China}

\author{Y.-M. Xu}
\affiliation{Materials Sciences Division, Lawrence Berkeley National Laboratory, Berkley, CA 94720, USA}

\author{X. Dai}
\affiliation{Beijing National Laboratory for Condensed Matter Physics, and Institute of Physics, Chinese Academy of Sciences, Beijing 100190, China}

\author{H. Ding}
\affiliation{Beijing National Laboratory for Condensed Matter Physics, and Institute of Physics, Chinese Academy of Sciences, Beijing 100190, China}\email{dingh@aphy.iphy.ac.cn}

\date{\today}% It is always \today, today,

\begin{abstract}
In order to improve the advantages and the reliability of the second derivative method in tracking the position of extrema from experimental curves, we develop a novel analysis method based on the mathematical concept of curvature. We derive the formulas for the curvature in one and two dimensions and demonstrate their applicability to simulated and experimental angle-resolved photoemission spectroscopy data. As compared to the second derivative, our new method improves the localization of the extrema and reduces the peak broadness for a better visualization on intensity image plots. 
\end{abstract}


\maketitle

%\tableofcontents

\section{Introduction}

With the development of multi-channel detectors and the recording of a huge amount of experimental data, the pass decade has witnessed  a boom in the use of color images for the representation of spectroscopic data in a very compact and easily visualized way. Typically, a color scale is associated with the experimental spectral intensity, which is displayed as a function of two independent variables. For example, such images are widely used in scanning tunneling microscopy (STM) [\onlinecite{Pan2001,Ma2008, Zhang2009,Hor2010}], Raman scattering [\onlinecite{Araujo2007, Kim2010, Krauss2010}], inelastic neutron scattering (INS) [\onlinecite{Zhao2009, Gilardi2004, Wakimoto2007, Doubble2010}], atomic force microscopy (AFM) [\onlinecite{Lauritsen2009,Liu_AFM2010, Kodera2010}], resonant inelastic X-ray scattering (RIXS) [\onlinecite{Abbamonte2004, Ishii2005,Smadici2009, Schlappa2009}] and angle-resolved photoemission spectroscopy (ARPES) [\onlinecite{Valla1999,Sato2001,Mesot2001,Ronning2003,Sato2005,Yang2005,Qian2006,Kim2006, Zhang_PRL2008, Richard2010, Liu_PRL2010, Borisenko2008}]. 

This imaging process is particularly efficient to represent energy band dispersions in the momentum or momentum-transfer spaces, where the energy and the momentum (or momentum-transfer) are the two independent variables. Frequently though, many bands or features overlap or have significant broadness, making direct visualization of the raw data difficult. The main tool commonly used in ARPES analysis to overcome this issue and improve direct visualization of band dispersion is the second derivative of intensity plots ~[\onlinecite{Sato2001,Mesot2001, Ronning2003, Sato2005, Yang2005, Qian2006, Kim2006, Zhang_PRL2008, Richard2010, Liu_PRL2010}]. Despite its success and widespread use, the method of second derivative gives sometimes results that differs slightly from the actual position of the maxima in the energy distribution curves (EDCs), where the photoemission intensity at fixed momentum is represented as a function of energy, or in the momentum distribution curves (MDCs), where the photoemission intensity at fixed energy is given as a function of momentum. Alternatives must thus be found to improve both accuracy and visualization of data.   

%With its ability to resolve directly the one-particle electronic spectra of crystals in energy and momentum ($k$), angle-resolved photoemission spectroscopy (ARPES) is one of the most powerful tool to access electronic band dispersion in materials. In addition to allow data analysis from energy distribution curves (EDC) and momentum distribution curves (MDC), the recent developments of multi-chanel detectors with a huge number of data point enables us to use color plots to literally visualize band dispersions in $k$-space ~[\onlinecite{Valla1999}]. Frequently however, many bands overlap or have significantly broadness, making direct visualization of the raw data difficult. The main tool commonly used to overcome this issue and improve direct visualization of band dispersion is the second derivative of intensity plots ~[\onlinecite{Sato2001,Mesot2001, Ronning2003, Sato2005, Yang2005, Qian2006, Kim2006, Richard2010}]. Despite its success and widespread use, the method of second derivative gives sometimes results that differs slightly from the actual position of the maxima of EDCs and MDCs. Alternatives must thus be found to improve both accuracy and visualization of data.

In this paper, we develop an analysis method for studying spectroscopic data based on the mathematical concept of curvature in one-dimension (1D) and two-dimension (2D). As an example, we apply this method to the study of electronic energy dispersion from ARPES data. We show two major advantages of the curvature method over the second derivative method: (i) the curvature method is more reliable in tracking the position of extrema and (ii) the curvature method can increase the sharpness of the dispersive features for a better visualization effect. We prove the efficiency of this method using both experimental and simulated data.

\section{1D curvature method}
The concept of curvature is used to quantitatively determine \emph{how much a curve is not straight}. It locally associates a radius of curvature, which can be either positive or negative, to a small segment along a curve. The mathematical definition in 1D of the curvature $C(\tilde{x})$ associated to a function $f(\tilde{x})$ is given by:

\begin{equation}\label{eq_1D}
C(\tilde{x}) = \frac{{f''(\tilde{x})}}{{(1 + f'(\tilde{x})^2 )^{\frac{3}{2}} }}
\end{equation}

\begin{figure*}[htbp]
\includegraphics[width=6.9in]{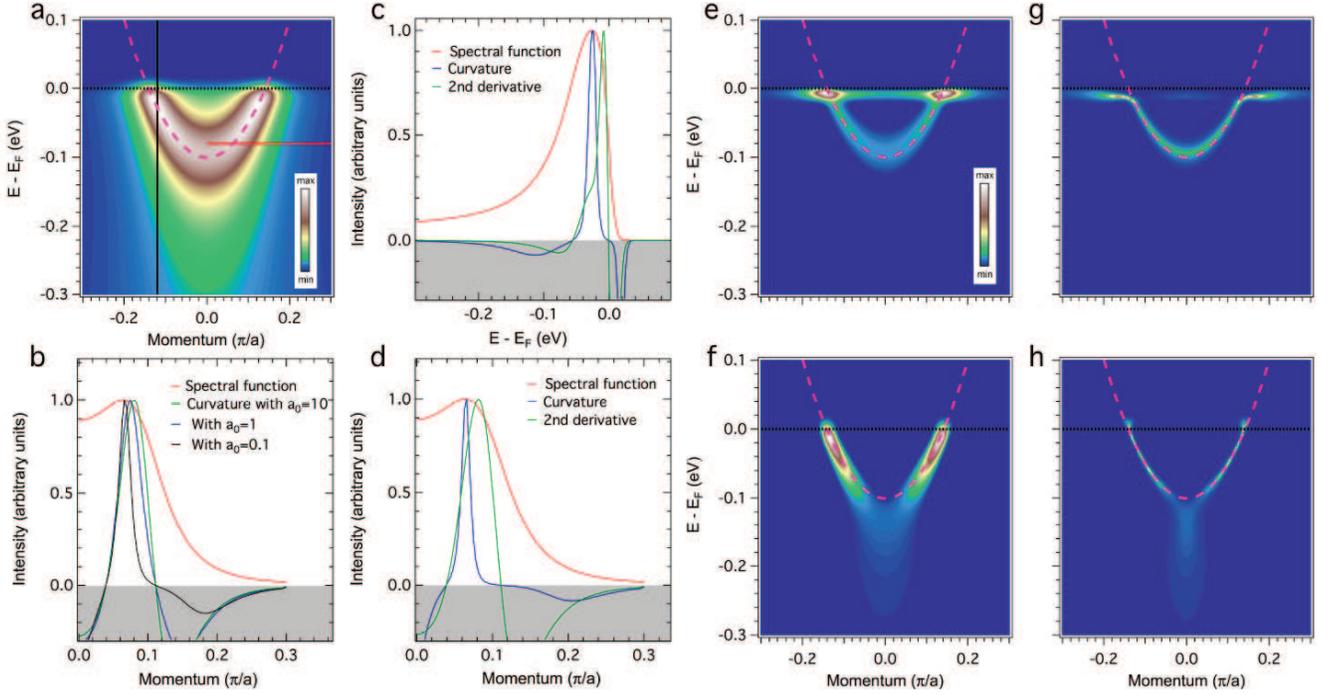}% Here is how to import EPS art
\caption{\label{fig:1Dcurv} Color online. (a) Simulated ARPES intensity plot (see the text). (b) Spectral intensity (red curve) as a function of momentum (MDC) along the horizontal red line in panel (a), compared to the curvature with different values of $a_0$ (see the text). (c) Comparison between the spectral intensity as a function of energy (EDC) along the vertical black line in panel (a), and the corresponding second derivative and curvature curves. (d) Same as (c) but for the spectral intensity as a function of momentum along the red line in panel (a). The intensity of each curve in panels (b)-(d) has been normalized to 1 and the sign of the second derivative and curvature curves has been reversed to facilitate visualization. The intensity plots of the second derivatives [curvature] of the simulated data from panel (a) along the energy and momentum directions are given in (e) [(g)] and (f) [(h)], respectively.}
\end{figure*}

For application to spectroscopic data, for example to an EDC curve, $f(\tilde{x})$ may represent the signal intensity whereas $\tilde{x}$ represents a unitless variable such as normalized energy. The normalization of a  variable $x$ that carries units is done through a transformation such as $x/\xi\rightarrow\tilde{x}$, where $\xi$ is a positive arbitrary constant with the same dimension as $x$. Since experimental spectroscopic functions themselves are usually defined to an arbitrary factor, $f(\tilde{x})$ carries the same information as $I_0f(\tilde{x})$, where $I_0$ is an arbitrary positive constant. Taking into account the arbitrariness in the absolute values of $\tilde{x}$ and $f(\tilde{x})$, we can rewrite equation (\ref{eq_1D}) as:    

\begin{equation}
C(x) = \frac{{{I_0\xi^2}f''(x)}}{{{{(1 + I_0^2\xi^2f'{{(x)}^2})}^{\frac{3}{2}}}}}
\end{equation}
Since we are interested uniquely in the relative variations of the curvature, this equation can be reduced further to:
\begin{equation}
C(x) \sim \frac{{f''(x)}}{{{{({C_0} + f'{{(x)}^2})}^{\frac{3}{2}}}}}
\end{equation}
where $C_0$ is a free parameter. In order to understand the meaning of $C_0$, we test the previous equation in two limit cases:

(1) When $C_0$ $\gg$ $f'(x)^2$, \emph{i.e}. when $f'(x)^2$ can be ignored, we get

\begin{equation}
C(x) \sim \frac{{f''(x)}}{{{{({C_0} + f'{{(x)}^2})}^{\frac{3}{2}}}}} \sim f''(x)
\end{equation}

\noindent which gives the same result as the second derivative method.

(2) When $C_0$ $\ll$ $f'(x)^2$,

\begin{equation}
C(x) \sim \frac{{f''(x)}}{{{{({C_0} + f'{{(x)}^2})}^{\frac{3}{2}}}}}\sim\frac{{f''(x)}}{{f'{{(x)}^3}}}
\end{equation}

This latter solution diverges at the extrema, where $f'(x)$ = 0. As $C_0$ approaches 0, the peak positions in $C(x)$ are getting closer and closer to the real peak positions. In the worst case, when $C_0\rightarrow\infty$, the curvature should provide a result as good as the one given by the second derivative. Therefore, the curvature is necessarily an improvement over the second derivative method in tracking the peak positions. In practice, we avoid singularities while maintaining the reliability of $C(x)$ by choosing an intermediate $C_0$. Empirically, we find out that the best compromise is reached when $C_0$ is of the order of the average or the maximum value of $|f'(x)|^2$. Hereafter, we express $C_0$ as $a_0|f'(x)|_{max}^2$, where $a_0$ is a positive constant and $|f'(x)|_{max}$ is the maximum value of $|f'(x)|$.

To illustrate the reliability of the curvature analysis, we simulate ARPES data using known parameters. The ARPES photoemission intensity can be expressed by the product of three terms: the Fermi-Dirac distribution $f_D(x)$, the spectral weight  $A({\bf k} ,\omega)$ that contains all the information about the dispersion, and a matrix element factor that depends on momentum, as well as on the energy and polarization of the probing photons. Since the latter term does not carry any information about the dispersion, we set it to 1. The spectral weight can be expressed in terms of the energy dispersion $\varepsilon_{\bf k} $ as:

\begin{equation}
A({\bf k} ,\omega ) =  - \frac{1}{\pi }\frac{{\Sigma ''({\bf k} ,\omega )}}{{(\omega  - \varepsilon _{\bf k}  - \Sigma '({\bf k} ,\omega ))^2  + \Sigma ''({\bf k} ,\omega )^2 }}
\end{equation}

\noindent where $\Sigma ({\bf k} ,\omega ){\rm{ = }}\Sigma '({\bf k} ,\omega ){\rm{ + i}}\Sigma ''({\bf k} ,\omega )$ is the self-energy of the quasi-particles. The self-energy is known to depend only weakly on momentum and its imaginary part usually varies like $\sim \alpha\omega^2+c$ at low energy. Thus, we set the self-energy to:

\begin{eqnarray}
\Sigma '(\omega ) =&  - \displaystyle\frac{{\alpha ((1 - c)\omega  - (1 + c){\omega ^3})}}{{\sqrt 2 (1 + {\omega ^4})}}\\
\Sigma ''(\omega ) =&  - \displaystyle\frac{{\alpha {\omega ^2} + c}}{{1 + {\omega ^4}}}
\end{eqnarray}

\noindent which satisfies the Kramers-Kronig transformation \cite{Hufner}.

Setting $\alpha=3$ and $c=0.15$ eV, we plot simulated ARPES data in Fig. \ref{fig:1Dcurv}(a) for the dispersion ${\varepsilon _{\bf k}}=15k^2-0.3$ eV at a temperature ($T$) of 20 K. The result has been further convoluted by a Gaussian function along the energy direction to simulate an energy resolution of 10 meV. In Fig.~\ref{fig:1Dcurv}(b), we compare the MDC along the red line in panel (a) to curvature curves of that same MDC using different values of $a_0$. For a better comparison, the sign of the curvature curves has been reversed and the maxima of all curves have been normalized to 1. As expected for an asymmetrical lineshape, the position of the curvature peak is slightly away from the real peak position when $a_0$ is large but converges to that latter position with $a_0$ decreasing. Moreover, the peak sharpens rapidly as $a_0$ decreases. Although this is obviously an advantage in tracking its position, we note that it is necessary to refrain decreasing $a_0$ too much while studying multi-feature systems since the sharpening of the peaks is accompanied by an increase of intensity in the curvature, which may affect the global contrast between all the features represented on a single image. We also note that since we are trying to find peak positions (maxima or inflections in the spectra), only the positive parts of the sign-reversed second derivatives and the sign-reversed curvatures have a physical meaning (the approximate position of peaks), and the negative parts are completely ignored. 

In Fig.~\ref{fig:1Dcurv}(c) and Fig.~\ref{fig:1Dcurv}(d), we plot the EDC and MDC along the black and red lines in panel (a), respectively, along with their second derivative and curvature curves (normalized and sign-reversed). Since both the MDC and EDC lineshapes are asymmetric with respect to the peak positions, the second derivative curves do not track the peak positions exactly and a small shift towards the highest slope change is observed. In contrast, the curvature analysis provides more reliable peak positions, in addition to giving sharper features. We performed the second derivative analysis for all EDCs and MDCs and we show the corresponding second derivative intensity plots in Fig.~\ref{fig:1Dcurv}(e) and Fig.~\ref{fig:1Dcurv}(f), respectively. Similarly, the EDC- and MDC-curvature intensity plots associated with the data of Fig.~\ref{fig:1Dcurv}(a) are given in Fig.~\ref{fig:1Dcurv}(g) and Fig.~\ref{fig:1Dcurv}(h), respectively. Obviously, the curvature method gives sharper features and allows a better tracking of the band dispersion as compared with the second derivative analysis. However, as for the analysis of  EDCs and MDCs and their corresponding second derivatives, the 1D curvature method presented here has some limitations over the whole range of energy and momentum. While the EDC-curvature method is quite reliable to track the minima and maxima of band dispersions, it gives unreliable results near the Fermi cutoff, which itself appears as a spectral feature. In contrast, the MDC-curvature method is quite precise near the Fermi cutoff but fails to reveal precisely the dispersion near extrema. Nevertheless, a cleaver combined use of EDC- and MDC-curvature analysis allows to track the band dispersion completely and precisely. A more sophisticated analysis method is proposed in the next section.   
 
\begin{figure}
\includegraphics[width=3.5in]{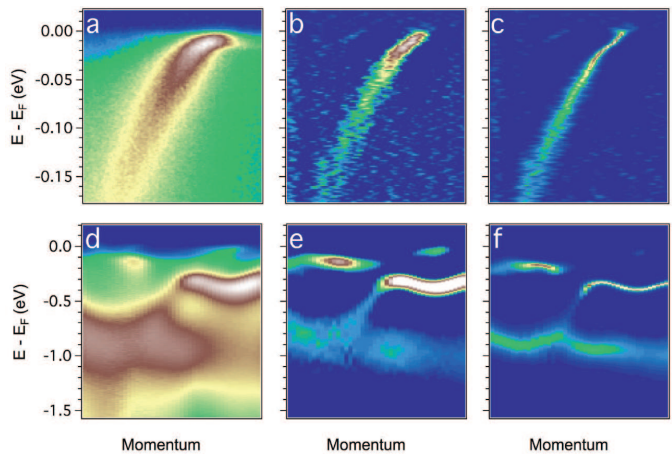}% Here is how to import EPS art
\caption{\label{fig:Appl} Color online. (a) ARPES intensity plot (from \cite{Ding_EPL2008}). (b)[(c)] Corresponding intensity plot of second derivative [1D curvature] along the momentum direction. (d) ARPES intensity plot (from \cite{Richard2009}). (e)[(f)] Corresponding intensity plot of second derivative [1D curvature] along the energy direction.}
\end{figure}

We now test the 1D curvature method on real experimental data. In Fig. \ref{fig:Appl}(a), we show an intensity plot recorded at 15 K corresponding to the low-energy band dispersion near the Fermi wavevector ($k_F$) of the so-called $\alpha$ band in optimally-doped Ba$_{0.6}$K$_{0.4}$Fe$_2$As$_2$ ($T_c$ = 37 K) [\onlinecite{Ding_EPL2008}]. As reported earlier, the dispersion exhibits in the superconducting state a kink or sudden slope change around 25 meV below the Fermi energy ($E_F$) due to an electron-mode coupling [\onlinecite{Richard2009}]. Although the kink is visible in the original image, it appears more clearly in the MDC-second derivative plot shown in Fig. \ref{fig:Appl}(b). As expected from the previous discussion, the result is even sharper with the use of the MDC-curvature method, as illustrated in Fig. \ref{fig:Appl}(c). The second derivative method is particularly efficient in ARPES for the study of band dispersion complexes. In Fig.~\ref{fig:Appl}(d), we show an ARPES intensity cut of Sr$_4$V$_2$O$_6$Fe$_2$As$_2$ recorded at 40 K along the $\Gamma-$M direction [\onlinecite{Qian2010}]. Within the wide energy range displayed (down to about 1.5 eV below $E_F$), many bands exist and overlap, and it is very difficult to extract their band dispersion. The corresponding EDC-second derivative intensity plot shown in Fig.~\ref{fig:Appl}(e) is a clear improvement for the visualization of the main bands. Once more, this advantage is reinforced with the EDC-curvature method, as illustrated in Fig.~\ref{fig:Appl}(f). The bands are sharper and the reliability in tracking the peak position improved. 

\section{2D curvature method}

Despite its ability to track band dispersions, the 1D curvature method has some unavoidable problems when analyzing intensity images. The main problem comes from the fact that the images themselves, as well as the features they emphasize, are 2D rather than 1D objects. In this section, we extend the 1D curvature method to a 2D method. As a first example, we treat the simplified case where the two independent variables determining the spectral intensity are equivalent. For example, this situation applies to AFM and STM mappings, for which both independent variables represent a distance, as well as to ARPES Fermi surface mappings, for which both independent variables represent a momentum component. Aftewards, we will focus on a more general case, where the independent variables are inequivalent, like in the energy \emph{vs} momentum intensity plots used in ARPES to reveal energy band dispersions. 

\subsection{Equivalent independent variables}

The equivalent in 2D of the second derivative is the Laplacian:

\begin{equation}
{\nabla ^2}f = \frac{{{\partial ^2}f}}{{\partial {\tilde{x}^2}}} + \frac{{{\partial ^2}f}}{{\partial {\tilde{y}^2}}}
\label{formula:lap}
\end{equation}

\noindent The passage from unitless variables $(\tilde{x},\tilde{y})$ to variables $({x},{y})$ with same units modifies the equation only by a global factor that does not affect the global contrast between different features on an image plot.  

Similarly to the second derivative, the mean curvature function has an equivalent in 2D for a function $f(\tilde{x},\tilde{y})$, which is given by:

\begin{equation}
\label{eq_curv2D}
C(\tilde{x},\tilde{y}) = \frac{{[1 + {{(\frac{{\partial f}}{{\partial \tilde{x}}})}^2}]\frac{{{\partial ^2}f}}{{\partial {\tilde{y}^2}}} - 2\frac{{\partial f}}{{\partial \tilde{x}}}\frac{{\partial f}}{{\partial \tilde{y}}}\frac{{{\partial ^2}f}}{{\partial \tilde{x}\partial \tilde{y}}} + [1 + {{(\frac{{\partial f}}{{\partial \tilde{y}}})}^2}]\frac{{{\partial ^2}f}}{{\partial {\tilde{x}^2}}}}}{{{{2[1 + {{(\frac{{\partial f}}{{\partial \tilde{x}}})}^2} + {{(\frac{{\partial f}}{{\partial \tilde{y}}})}^2}]}^{\frac{3}{2}}}}}
\end{equation}

\begin{figure}[htbp]
\includegraphics[width=8cm]{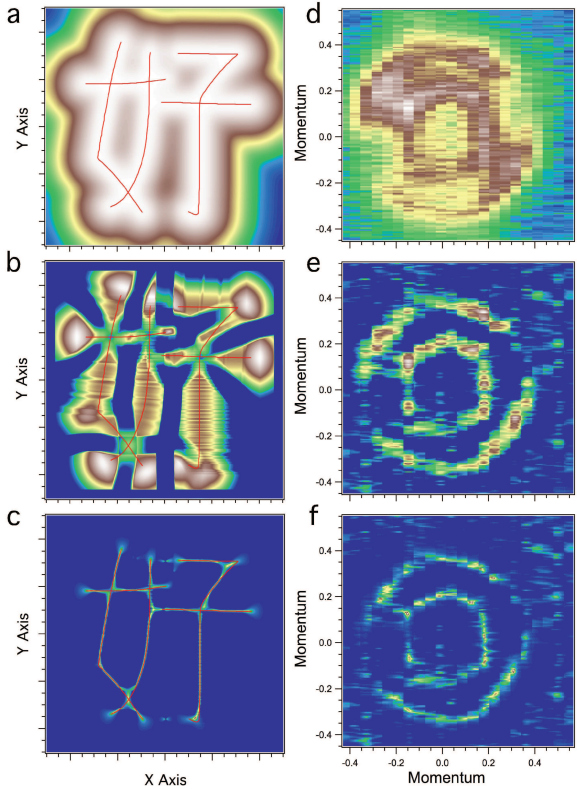}% Here is how to import EPS art
\caption{\label{fig:2Dcurv} Color online. (a) Image representation of the Chinese character \emph{h\v{a}o} (see the text). (b)[(c)] Corresponding intensity plot of the Laplacian [2D curvature]. The original character in (a)-(c) is given by the red lines. (d) ARPES Fermi surface mapping of Ba$_{0.6}$K$_{0.4}$Fe$_2$As$_2$. (e)[(f)] Corresponding intensity plot of the Laplacian [2D curvature].}
\end{figure}

When the independent variables carry the same units, we need to use the transformations $\frac{{\partial}}{{\partial \tilde{x}}} \to \xi\frac{{\partial}}{{\partial x}}$ and $\frac{{\partial}}{{\partial \tilde{y}}} \to \xi\frac{{\partial}}{{\partial y}}$. Considering that the spectral function $f$ is defined to a factor $I_0$, we get:

\begin{widetext}
\begin{equation}
C(x,y) \sim \frac{{[C_0 + {{(\frac{{\partial f}}{{\partial x}})}^2}]\frac{{{\partial ^2}f}}{{\partial {y^2}}} - 2\frac{{\partial f}}{{\partial x}}\frac{{\partial f}}{{\partial y}}\frac{{{\partial ^2}f}}{{\partial x\partial y}} + [C_0 + {{(\frac{{\partial f}}{{\partial y}})}^2}]\frac{{{\partial ^2}f}}{{\partial {x^2}}}}}{{{{[C_0 + {{(\frac{{\partial f}}{{\partial x}})}^2} + {{(\frac{{\partial f}}{{\partial y}})}^2}]}^{\frac{3}{2}}}}}
\end{equation}
\end{widetext}

\noindent where a global factor has been removed and $C_0=(I_0\xi)^{-2}$ is a free positive parameter.

Let's now compare both 2D methods. In Fig. \ref{fig:2Dcurv}(a), we plot a Chinese character (\emph{h\v{a}o}, which means ``good"). The character has been broaden by a Gaussian distribution and further blurred by a boxcar filter. Although the character is recognizable on the raw image, the strokes are not sharp. The Laplacian of this image is displayed in Fig. \ref{fig:2Dcurv}(b). While the Laplacian allows to sharpen the strokes a little, the latter remain broad and the whole character appears distorted. In contrast, the result obtained by the 2D curvature method and shown in Fig. \ref{fig:2Dcurv}(c) gives a much better representation of the original character, with very sharp strokes. Only little distortion can be observed near stroke intersections and near the beginning and the end of each stroke. Analysis of real ARPES data with experimental noise leads to similar conclusion. In Fig. \ref{fig:2Dcurv}(d), we display the ARPES photoemmission intensity mapping around the Brillouin zone center of a Ba$_{0.6}$K$_{0.4}$Fe$_2$As$_2$ sample, which has been integrated over a $\pm$ 10 meV energy range around the Fermi level. The high intensity regions represent the Fermi surface. Although the raw data are sufficient to distinguish the presence of two Fermi surface sheets [\onlinecite{Ding_EPL2008}], the Fermi surface contours are difficult to identify precisely. In this case, the Laplacian improves the Fermi surface determination of the two concentric Fermi surfaces centered at the Brillouin zone center. Further improvement is provided by the 2D curvature, which makes the Fermi surface contours narrower. 

\subsection{Inequivalent independent variables}

Unfortunately, spectroscopic data cannot always be presented as 2D mappings with $x$ and $y$ axes having the same units. This is particularly true when dealing with the momentum space, like in ARPES, INS and RIXS. Commonly, the results may represent the spectral intensity as a function of energy, and momentum or momentum-transfer. In that case, the Laplacian can be adapted to variables $x$ and $y$ with different units by using the transformations $\frac{{\partial}}{{\partial \tilde{x}}} \to \xi\frac{{\partial}}{{\partial x}}$ and $\frac{{\partial}}{{\partial \tilde{y}}} \to \eta\frac{{\partial}}{{\partial y}}$, where $\xi$ and $\eta$ are positive parameters carrying the same units as $x$ and $y$, respectively. Accounting once more for a global positive factor $I_0$ in the absolute value of the experimental spectral response $f$, we obtain:

\begin{eqnarray}
\label{Laplace_units}
{\nabla ^2}f = &I_0\xi^2\frac{{{\partial ^2}f}}{{\partial {x^2}}} + I_0\eta^2\frac{{{\partial ^2}f}}{{\partial {y^2}}}\\
\sim &(\frac{\xi}{\eta})^2\frac{{{\partial ^2}f}}{{\partial {x^2}}} + \frac{{{\partial ^2}f}}{{\partial {y^2}}}
\end{eqnarray}

\noindent Where we removed a global factor. The latest equation has only one independent parameter, $\xi/\eta$. A natural choice of parameter to capture the main features in an image plot is to make the second derivative terms of the same order of magnitude, which is done by setting the ranges of the data in $x$ and $y$ to similar values. For a square grid for example (same number of columns and rows), that statement is equivalent to $\xi/\eta=\Delta x/\Delta y$, where $\Delta x$ and $\Delta y$ are the stepsizes along the $x$ and $y$ axes, respectively. 

Similarly to the Laplacian, equation \eqref{eq_curv2D} can be adapted to variables $x$ and $y$ with different units. Using the same transformations for $\tilde{x}$ and $\tilde{y}$, we get:  

\begin{widetext}
\begin{equation}
\label{eq_inequivalent}
C(x,y) \sim \frac{{[1 + C_x{{(\frac{{\partial f}}{{\partial x}})}^2}]C_y\frac{{{\partial ^2}f}}{{\partial {y^2}}} - 2C_xC_y\frac{{\partial f}}{{\partial x}}\frac{{\partial f}}{{\partial y}}\frac{{{\partial ^2}f}}{{\partial x\partial y}} + [1+ C_y{{(\frac{{\partial f}}{{\partial y}})}^2}]C_x\frac{{{\partial ^2}f}}{{\partial {x^2}}}}}{{{{[1 + C_x{{(\frac{{\partial f}}{{\partial x}})}^2} + C_y{{(\frac{{\partial f}}{{\partial y}})}^2}]}^{\frac{3}{2}}}}}
\end{equation}
\end{widetext}

\noindent where $C_x=I^2_0\xi^2$ and $C_y=I^2_0\eta^2$ are the only two (positive) free parameters for this equation. Using the same arguments as for the Laplacian, we can set $\xi/\eta=\Delta x/\Delta y$ to assure a good visual representation. In this condition, we verify easily that in the limit where $I_0\to 0$, and thus $C_x\to 0$ and $C_y\to 0$, equation \eqref{eq_inequivalent} is simplified to

\begin{equation}
C(x,y) \sim C_x\frac{{{\partial ^2}f}}{{\partial {x^2}}} + C_y\frac{{{\partial ^2}f}}{{\partial {y^2}}}
\end{equation}

\noindent which is equivalent to our definition given in equation \eqref{Laplace_units} of the Laplacian with variables carrying units. In the opposite limit, when $I_0\to\infty$, we find:

%\begin{widetext}
\begin{equation}
C(x,y) \sim \frac{{{{(\frac{{\partial f}}{{\partial x}})}^2}\frac{{{\partial ^2}f}}{{\partial {y^2}}} - 2\frac{{\partial f}}{{\partial x}}\frac{{\partial f}}{{\partial y}}\frac{{{\partial ^2}f}}{{\partial x\partial y}} + {{(\frac{{\partial f}}{{\partial y}})}^2}\frac{{{\partial ^2}f}}{{\partial {x^2}}}}}{{{{[ C_x{{(\frac{{\partial f}}{{\partial x}})}^2} + C_y{{(\frac{{\partial f}}{{\partial y}})}^2}]}^{\frac{3}{2}}}}}
\end{equation}
%\end{widetext}

\noindent The latest equation diverges when:

\begin{eqnarray}
[C_x(\frac{\partial f}{\partial x})^2+C_y(\frac{\partial f}{\partial y})^2]^{\frac{3}{2}}=0\\
\Rightarrow |\nabla f(\tilde{x},\tilde{y})| =0
\end{eqnarray}

\noindent which corresponds exactly to the position of the extrema of $f$. Therefore, we conclude that the 2D curvature is necessarily an improvement compared to the Laplacian in tracking the position of extrema.

\begin{figure}[htbp]
\includegraphics[width=8cm]{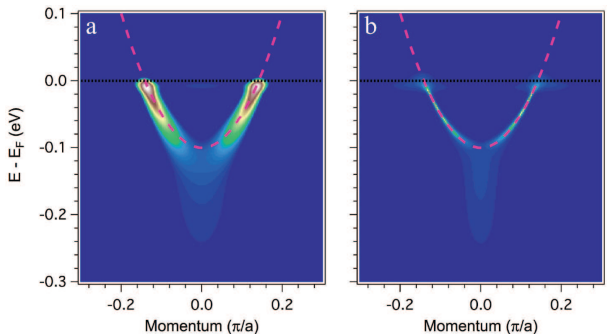}% Here is how to import EPS art
\caption{\label{2Ddispersion} Color online. (a) Laplacian of the simulated ARPES intensity plot shown in Figure 1(a). (b) 2D curvature of the simulated ARPES intensity plot shown in Figure 1(a).}
\end{figure}

In Figure \ref{2Ddispersion}, we compare the Laplacian and the 2D curvature intensity plots for the simulated electronic dispersion given in Figure \ref{fig:1Dcurv}(a). As expected, the 2D curvature method gives sharper features. In addition, it tracks the original band dispersion with higher accuracy over the whole range of energy. It is also instructive to note that while the 1D curvature method using EDCs and MDCs gives results better than the 2D curvature near the band bottom and near the Fermi level, respectively, the 2D curvature is more reliable over the whole energy range.  

\section{Discussion}

As with the second derivative method, the curvature analysis technique described in this paper is a powerful method to enhance dispersive features in a spectroscopic image. It is very important to keep in mind that this is its only purpose and that the information contained in the original spectra is indeed richer, despite being sometimes difficult to access. These visualization methods can thus be regarded as effective complementary tools in understanding spectroscopic data. For example, while the precise shape of MDCs and EDCs from ARPES data are often intimately related to intrinsic scattering and other electronic interactions, information completely lost in the curvature intensity plots, MDCs and EDCs are not always good ways to represent dispersion. This is especially true for multi-bands systems when bands are broad. Besides, band dispersions are 2D objects ($k$ \emph{vs} $E$), which are thus more naturally represented by a 2D image plot. Indeed, MDC- and EDC-analysis in ARPES often lead to slightly different dispersion, even though real electronic dispersions, namely $E$ \emph{vs} $k$ relationships, are uniquely defined objects. By using the 2D curvature method described here, it is possible to remove this ambiguity. However, we note that such analysis is accurate only when we dispose of sufficient data along both directions ($E$ and $k$).    

Although the curvature technique constitutes an obvious improvement over the second derivative method in terms of reliability and sharpness of the spectral features, its main apparent disadvantage is the introduction of arbitrary parameters. As shown above, the curvature method is at least as reliable as the second derivative method in tracking the peak position of dispersive features, whatsoever the parameters used. Similarly, the sharpness of the dispersive features is also improved compared to the second derivative method. In that sense, the arbitrariness of the parameters is not a handicap. In fact, it gives some latitude to tune the relative contrast between different features from a single image and allow a better visualization effect.

\section{Conclusions}
We have developed a method based on the concept of curvature to analyze spectroscopic image plots. As with the second derivative method, which is widely used, the method presented here is quite efficient for representing dispersive features. Using simulated and experimental spectral images, we demonstrated that compared to second derivative analysis, the new curvature method improves significantly the reliability in tracking dispersive feature. Moreover, it sharpens spectral features for a better visualization of the spectroscopic features. 

\begin{acknowledgments}
We acknowledge useful discussions with Y. B. Huang, X. P. Wang, T. J. Min, T. Ayral and A. Van Roekeghem. This work was supported by grants from CAS (2010Y1JB6), NSFC (11004232 and 11050110422) and MOST of China (2010CB923000). 
\end{acknowledgments}

%\bibliography{main}% Produces the bibliography via BibTeX.

\end{document}